\documentclass[12pt, twoside]{article}
\usepackage{amstext,amsfonts}

\begin{document}

\vspace*{-1cm}
\begin{flushright}
IFIC--01--37
\\
\end{flushright}

\begin{center}
\begin{Large}
\bfseries{Kaluza-Klein theory, AdS/CFT correspondence and black hole entropy}
\end{Large}

\vspace*{1.6cm}

\begin{large}
J.M. Izquierdo$^{a1}$, J. Navarro-Salas$^{b1}$ and P. Navarro$^{b}$
\footnote{e-mails: izquierd@fta.uva.es,
jnavarro@hal.ific.uv.es, pnavarro@lie.ific.uv.es}
\end{large}
\vspace*{0.6cm}

\begin{it}
$a$ Departamento de F\'{\i}sica Te\'orica, Universidad de Valladolid
\\
E--47011 Valladolid, Spain
\\[0.4cm]
$b$ Departamento de F\'{\i}sica Te\'orica, Universidad de Valencia
\\
and IFIC, Centro Mixto Universidad de Valencia--CSIC
\\
E--46100 Burjassot (Valencia), Spain
\end{it}
\vspace*{1.6cm}

\end{center}
\vspace*{1.4cm}

\begin{abstract}
The asymptotic symmetries of the near-horizon geometry of a lifted 
(near-extremal) Reissner-Nordstrom black hole, obtained by inverting
the Kaluza-Klein reduction, explain the deviation of the 
Bekenstein-Hawking entropy from extremality. We point out the fact that the 
extra dimension allows us to justify the use of a Virasoro mode decomposition
along the time-like boundary of the near-horizon geometry, AdS$_2\times$S$^n$,
of the  lower-dimensional (Reissner-Nordstrom) spacetime.
\end{abstract}

\newpage

\section{Introduction}

The universality of the Bekenstein-Hawking area law of black holes could be
explained if the density of the microscopic states is controlled by the
conformal symmetry \cite{C00}. A nice example of this philosophy is 
provided by the BTZ black holes \cite{BTZ92}. The Bekenstein-Hawking
entropy can be derived \cite{S98}, via Cardy's formula, from the two-dimensional
conformal symmetry arising at spatial infinity of three-dimensional
gravity with a negative cosmological constant \cite{BH86}. Moreover,
one can look directly at the black hole horizon and treat it as a boundary.
The constraint algebra of surface deformations of the boundary, with 
additional ad-hoc conditions leads to a Virasoro algebra with a non-vanishing 
central charge \cite{S99,C99a,C99b}. With the aid of the Cardy formula it
is possible to reproduce the adequate density of states leading to the
Bekenstein-Hawking area law. Both approaches can be considered 
simultaneously if one looks at the asymptotic boundary of the near-horizon
geometry of black holes. This way one can also explain the entropy of
higher -dimensional black holes arising in string theory whose near-horizon
geometries are similar to the three-dimensional BTZ black holes.

However, in relevant black holes of general relativity (such as 
the Reissner-Nordstrom 
solutions) AdS$_3$ does not appear in the near-horizon geometry. It contains,
instead, AdS$_2$. The asymptotic symmetries of the AdS$_2$ metric generate
a Virasoro algebra \cite{CM99} with a calculable central charge capable to
reproduce the near-extremal black hole entropy \cite{NN00,NNN00}. These
Virasoro symmetries are those ($t$--$r$)-diffeomorphisms which leave
invariant the large $r$ behaviour of the AdS$_2$ metric. Therefore the Virasoro
generators can be regarded as living in the boundary of AdS$_2$, and to perform
a mode decomposition one has to make an integration in the time direction.
But the canonical formalism requires integration over a space-like slice.
So to properly study the $d=2$ case of the AdS/CFT correspondence
\cite{AGMOO00}, in terms of
asymptotic symmetries, it is necessary to clarify this question. The 
problem of this extra time-integration also appears in the approach of 
\cite{C99b}, where it is introduced to properly obtain a central term for a 
Virasoro algebra. The main aim of this paper is to improve the understanding
of this issue: Why do the Virasoro modes in the unconventional time-direction
produce the adequate central charge to explain the black hole entropy? We 
shall focus our attention in the Reissner-Nordstrom  (R-N) black holes (in any
dimension) and find a natural answer in the context of the Kaluza-Klein (KK)
theory.

In Section 2 we shall show how the near-horizon configuration of the
Reissner-Nordstrom solution can be lifted, inverting the Kaluza-Klein
decomposition, to a higher-dimensional metric containing AdS$_3$. In
Section 3 we then analyze the AdS$_3$/CFT$_2$ and AdS$_2$/CFT$_1$ 
correspondences within this context, showing that, irrespective of 
the particular realization of the asymptotic symmetries, the value of the
central charge is unique. In Section 4 we argue, from the higher-dimensional
point of view, that the mode decomposition in the time direction can be
justified. When we add the extra dimension the ordinary time coordinate
is converted into a null one in the boundary of the higher-dimensional
theory. Therefore one can use Cardy's formula to evaluate, as in \cite{S98}, the
asymptotic growth of states reproducing then the Beckenstein-Hawking 
formula around extremality. Finally, in Section 5, we summarize our 
conclusions.

\section{Relating near-extremal R-N and BTZ black holes by
inverting the KK mechanism}

Let us consider a R-N black hole configuration in $(n+2)$ dimensions. The metric 
field is
\begin{equation}
  ds^2_{(n+2)}=-U(r)dt^2+\frac{dr^2}{U(r)}+r^2d\Omega^2_{(n)}\quad ,
                                                              \label{2.1}
\end{equation}
where
\begin{equation}
      U(r)=1-\frac{2G_{n+2}M}{\Gamma_{(n)}r^{n-1}}+
    \frac{G_{n+2}^2Q^2}{r^{2(n-1)}\Delta_{(n)}}\quad ,       \label{2.2}
\end{equation}
\begin{equation}
      \Gamma_{(n)}=\frac{n\nu^{(n)}}{8\pi}\ ,\quad \Delta_{(n)}=
   \frac{n}{2(n-1)}\quad ,                                    \label{2.3}
\end{equation}
$\nu^{(n)}$ is the area of the unit $S^n$ sphere,
\begin{equation}
   \nu^{(n)}=\frac{n\pi^{(n+1)/n}}{\Gamma\left(\frac{n+1}{n}\right)}\label{2.4}
\end{equation}
and the electromagnetic field is given by
\begin{equation}
     A=G_{n+2}\frac{Q}{r^{n-1}}dt\quad .                       \label{2.5}
\end{equation}
$Q$ is the electric charge and $G_{n+2}$ is the $(n+2)$-dimensional Newton's
constant in geometrized units.

When the R-N configuration is close to extremality and we consider the
near-horizon region as follows ($0<\alpha <<1$)
\begin{eqnarray}
     M &=& M_0(1+k\alpha^2)             \label{2.6}\\
    r &=& r_0+\alpha x \quad ,              \label{2.7}
\end{eqnarray}
where $M_0$ and $r_0$ are the extremal mass and radius
\begin{equation}
    M_0=\frac{|Q|\Gamma_{(n)}}{\Delta^{1/2}_{(n)}}\quad ,       \label{2.8}
\end{equation}
\begin{equation}
      r_0=\left(\frac{G_{n+2}M_0}{\Gamma_{(n)}}\right)^{\frac{1}{n-1}}
  \quad ,                                                     \label{2.9}
\end{equation}
the leading terms in a series expansion in $\alpha$ of (\ref{2.1}) and
(\ref{2.5}) are
\begin{equation}
   ds^2_{(n+2)}=-\left( (n-1)^2\frac{\alpha^2x^2}{r_0^2}-2\frac{\Delta M}{M_0}
\right) dt^2+\frac{\alpha^2dx^2}{\frac{(n-1)^2\alpha^2x^2}{r_0^2}-2
\frac{\Delta M}{M_0}}+r_0^2d\Omega_{(n)}^2\ ,                 \label{2.10}
\end{equation}
where $\Delta M$ is de deviation of the mass from extremality, $\Delta M=
M-M_0$, and
\begin{equation}
   A=\frac{QG_{n+2}}{r_0^{n-1}}\left( 1-(n-1)\frac{\alpha x}{r_0}\right)dt
\quad .                                                    \label{2.11}
\end{equation}
The metric (\ref{2.10}) is just the Robinson-Bertotti geometry 
AdS$_2\times$S$^n$, with two-dimensional curvature (in the $t$--$r$ plane) $R^{(2)}=
-\frac{2(n-1)^2}{r_0^2}$.

The idea now is to construct a higher-dimensional geometry from the metric and
the gauge field by inverting the KK reduction. So the new metric is
\begin{equation}
   ds^2_{(n+3)}=ds^2_{(n+2)}+\mu^2(d\theta+QA_\mu dx^\mu)^2\quad ,  \label{2.12}
\end{equation}
where the KK radius $\mu$ is given by
\begin{equation}
  \mu^2=\frac{1}{\Delta_{(n)}Q^2}\quad .                     \label{2.13}
\end{equation}
At leading order in $\alpha$ we have
\begin{eqnarray}
    ds^2_{(n+3)} &=& -\left( \frac{(n-1)^2\alpha^2x^2}{r_0^2} 
-2\frac{\Delta M}{M_0}\right) dt^2 +
\frac{\alpha^2dx^2}{\frac{(n-1)^2\alpha^2x^2}{r_0^2}-2
\frac{\Delta M}{M_0}}+r_0^2d\Omega_{(n)}^2\nonumber\\
 &+& \mu^2\left[ d\theta+\frac{Q^2G_{n+2}}{r_0^{n-1}}\left(1-(n-1)
\frac{\alpha x}{r_0}\right)dt\right]^2                    \quad . \label{2.14}
\end{eqnarray}
The relevant fact is that the three-dimensional ($t,r,\theta$) part of the
above metric is just a BTZ black hole with curvature
\begin{equation}
     R^{(3)}=-\frac{2}{l^2}                                   \label{2.15}
\end{equation}
where
\begin{equation}
    l^2=\frac{4r_0^2}{(n-1)^2}                            \label{2.16}
\end{equation}

To explicitly relate the three-dimensional sector of (\ref{2.14}) with the
BTZ black hole we have to perform the natural identification
\begin{equation}
      \theta+\frac{Q^2}{r_0^{n-1}}G_{n+2}t=x^+\quad ,      \label{2.17}
\end{equation}
together with
\begin{equation}
     2|Q|\Delta_{(n)}^{1/2}t=x^-\quad ,                  \label{2.18}
\end{equation}
and
\begin{equation}
   \bar r^2=\frac{(n-1)}{Q^2\Delta_{(n)}r_0}\alpha x \quad .\label{2.19}
\end{equation}
Plugging these changes of coordinates into (\ref{2.14}) we get the 
following three-dimensional metric
\begin{equation}
    ds^2_{(3)}=-\bar r^2 dx^+dx^- +\gamma_{--}(dx^-)^2+\gamma_{++}
  (dx^+)^2+\frac{l^2\bar r^2d\bar r^2}{\bar r^4-4\gamma_{++}\gamma_{--}}
                                                              \label{2.20}
\end{equation}
where
\begin{eqnarray}
      \gamma_{--} &=& \frac{G_{n+2}}{2r_0^{n-1}\Gamma_{(n)}Q^2\Delta_{(n)}}
 \Delta M\quad ,                                              \label{2.21}\\
\gamma_{++} &=& \frac{1}{\Delta_{(n)}Q^2}\quad .             \label{2.22}
\end{eqnarray}
It is interesting to remark that for general chiral functions 
$\gamma_{++}(x^+)$, $\gamma_{--}(x^-)$ we have the general solution of 
three-dimensional gravity with a negative cosmological constant 
$-\frac{2}{l^2}$. For constant values of $\gamma_{\pm\pm}$ we have BTZ
black holes with 
\begin{equation}
  \gamma_{\pm\pm}=2G_3l(M_{BTZ}l\pm J)\quad ,             \label{2.23}
\end{equation}
where $M_{BTZ}$ and $J$ are the corresponding mass and angular momentum.

If we could establish a one-to-one correspondence between the near-extremal
R-N and BTZ black holes we could compare the corresponding entropies. There
is a straighforward way to do this by comparing the different actions
obtained by dimensional reduction. Starting
from the $(n+3)$-dimensional action
\begin{equation}
     \frac{1}{16\pi G_{n+3}}\int d^{n+3}x\sqrt{-g}R\quad , \label{2.24}
\end{equation}
the KK reduction with constant radius (\ref{2.13}) implies that
\begin{equation}
     \frac{1}{G_{n+2}}=\frac{1}{G_{n+3}}2\pi\mu\quad ,  \label{2.25}
\end{equation}
and the factorization of the angular coordinates in the near-horizon 
region gives
\begin{equation}
   \frac{1}{G_3}=\frac{1}{G_{n+3}}\nu^{(n)}r_0^n\quad .   \label{2.26}
\end{equation}
Using the above expressions one finds that
\begin{equation}
   G_3=G_{n+2}\frac{2\pi\mu}{\nu^{(n)}r_0^n}\quad .       \label{2.27}
\end{equation}

The entropy formula for tha BTZ black holes is 
\begin{equation}
 S_{BTZ}=\pi\sqrt{\frac{l(lM_{BTZ}+J)}{2G_3}}+
     \pi\sqrt{\frac{l(lM_{BTZ}-J)}{2G_3}}                \label{2.28}
\end{equation}
whereas the formula for R-N black holes is
\begin{equation}
   S_{RN}=\frac{\nu^{(n)}r_+^n}{4G_{n+2}}\quad ,          \label{2.29}
\end{equation}
where $r_+$ is the outer horizon. Expanding (\ref{2.29}) around extremality
we find
\begin{equation}
    S_{RN}=\frac{\nu^{(n)}r_0^n}{4G_{n+2}}+2\pi\sqrt{\frac{2r_0^2M_0}{(n-1)^2}
   \Delta M}   \quad .                                     \label{2.30}
\end{equation}
Comparing (\ref{2.28}) and (\ref{2.30}) using the relations 
(\ref{2.21})-(\ref{2.23}) and (\ref{2.27}) we find agreement only in the 
extremal configurations $\Delta M=0$ and $M_{BTZ}l=|J|$. We then
obtain a result similar to that found in \cite{S00}. In this reference
it was shown that a R-N black hole can be regarded as a dimensional 
reduction of a boosted black string and the near-horizon of the latter
contains a BTZ black hole. The entropies of the R-N black hole and the
associated BTZ black hole agree at extremality. We have carried out a 
similar argument by replacing the black string construction by the KK
decomposition. However the above mechanisms fail when one compares
the entropy deviations from extremality. To properly account
for it we shall follow a different 
route based on the symmetry properties of the near-horizon geometries. This
is the main aim of this paper and it will be considered in the next sections.

\section{Asymptotic symmetries and boundary CFT}

The Bekenstein-Hawking area law of BTZ black holes can be understood in
terms of the asymptotic symmetries of (\ref{2.20}) \cite{S98}. The infinitesimal
diffeomorphisms $\zeta^a$ preserving the asymptotic $r\rightarrow\infty$
form of (\ref{2.20}) are
\begin{eqnarray}
   \zeta^{\pm} &=& \varepsilon^\pm(x^\pm)+\frac{l^2}{2r^2}\partial^2_{\mp}
  \varepsilon^\mp(x^\mp)+\dots                   \label{3.1}\\
    \zeta^{\bar r} &=& -\frac{\bar r}{2}(\partial_+\varepsilon^+ +
    \partial_-\varepsilon^-)+\dots                 \label{3.2}   
\end{eqnarray}
where $\varepsilon^\pm(x^\pm)$ are arbitrary chiral functions. Therefore it
seems natural to investigate whether these sort of symmetries can also
explain the black hole entropy of the near-extremal R-N black holes. The
above transformations are asymptotic symmetries of the higher-dimensional
near-horizon theory. However only the transformations generated by
$\varepsilon^-(x^-)$ preserve the KK decomposition and can be pushed down. 
We must note that, in the R-N space-time, the coordinate $x^-$ is just a
time-like coordinate (see (\ref{2.18})), and therefore the transformations
generated by $\varepsilon^-(x^-)$ can be rewritten as
\begin{eqnarray}
   \zeta^t &=& \varepsilon(t)+\dots               \label{3.3}\\
    \zeta^x &=& -x\partial_t\varepsilon(t)+\dots   \label{3.4}\\
    \zeta^+ &=& \frac{l^2Q^2r_0}{2(n-1)\alpha x}\frac{1}{2|Q|
    \Delta_{(n)}^{1/2}}\partial^2_t\varepsilon(t)+\dots \quad . \label{3.5}
\end{eqnarray}
The infinitesimal transformations (\ref{3.3}), (\ref{3.4}) are diffeomorphisms
in the R-N spacetime whereas (\ref{3.5}) can be interpreted, as usual, as
a gauge transformation. However these transformations do not preserve the 
asymptotic form of the two-dimensional anti-de Sitter metric, 
although they preserve the asymptotic form of the gauge field. To
maintain simultaneously the asymptotic behaviour of the two-dimensional
metric and gauge field one can alternatively introduce the following
transformations:
\begin{eqnarray}
  \zeta^t &=& \varepsilon(t)+\frac{l^4}{32x^2\alpha^2}\varepsilon{''}(t)+\dots
                                                   \label{3.6}\\
  \zeta^x &=& -x\varepsilon{'}(t)+\dots         \label{3.7}\\
   \zeta^+ &=& 0 \quad .                         \label{3.8}
\end{eqnarray}
In contrast with (\ref{3.3})-(\ref{3.5}), they are pure diffeomorphisms in the R-N 
spacetime and the gauge transformation is trivial.
We have to remark that these transformations do not preserve the asymptotic
form of the three-dimensional anti-De Sitter metric (\ref{2.20}). However,
both sets of transformations can be used to provide realizations of the 
AdS/CFT correspondence in
a consistent way. In the first case it will be a chiral sector of the
AdS$_3$/CFT$_2$ correspondence, whereas in the second case we shall deal 
with a sort of 
AdS$_2$/CFT$_1$ correspondence. Our task now is to analyze these
two sets of symmetries on the spacetime parametrized by $\{ t,x\}$ and
determine the relevant quantities of the boundary theory required to
compute the entropy via Cardy's formula. To this end we shall use the 
technique introduced in \cite{NN98},
and assume the following asymptotic behaviour of the two-dimensional
metric and gauge field
\begin{eqnarray}
    g_{tt} &=& -\frac{4}{l^2}\alpha^2x^2+\gamma_{tt}+\dots \label{3.9}\\
 g_{tx} &=& \frac{\gamma_{tx}}{x^3}+\dots \label{3.10}\\
  g_{xx} &=& \frac{l^2}{4x^2}+\frac{\gamma_{xx}}{x^4}+\dots\label{3.11}\\
 A_t &=& \frac{Q}{|Q|}\Delta_{(n)}^{1/2} \left[ 1-(n-1)\left(\frac{\alpha x}{r_0}
  -\frac{r_0}{2\alpha x}\gamma_{A_t}+\dots \right)\right] \label{3.12}\\
  A_x &=& \frac{\gamma_{A_x}}{x^3}+\dots  \quad .            \label{3.13}
\end{eqnarray}
The action on the radial function, which sometimes is considered as a dilaton,
plays no role here. In fact, it could be considered constant. It is just the
gauge field which plays a fundamental role.

\subsection{AdS$_3$-symmetries}

We shall now consider the 
transformations (\ref{3.3})-(\ref{3.5}), which preserve the asymptotic form
of the lifted three-dimensional metric. One can show that the unique 
quantity invariant under the ``pure gauge transformations"
\begin{eqnarray}
   \zeta^t &=& \frac{\alpha^t(t)}{x^2}        \label{3.14} \\
   \zeta^x &=& \alpha^x(t)                  \label{3.15}\\
   \zeta^+ &=&  \frac{\alpha^+(t)}{x^2}              \label{3.16}
\end{eqnarray}
is given by
\begin{equation}
      \Theta_{tt}=k_1(\gamma_{tt}-(n-1)^2\gamma_{A_t})     \label{3.17}
\end{equation}
where $k_1$ is a constant coefficient. Moreover the transformation law of
$\Theta_{tt}$ is
\begin{equation}
   \delta_{\varepsilon(t)}\Theta_{tt}=\varepsilon(t)\Theta{'}_{tt}+
2\Theta_{tt}\varepsilon{'}(t)-\frac{k_1l^2}{2}\varepsilon{'''}(t)
      \quad ,                                             \label{3.18}
\end{equation}
which coincides with that of a chiral component of the stress tensor of a
two-dimensional conformal field theory.
The easiest way to determine $k_1$, and hence the central charge, is to
realize that the value of $\Theta_{tt}$, for a static R-N black hole, should
coincide with the mass. More properly, the deviation of the mass from 
extremality
\begin{equation}
   \Theta_{tt}\vert_{RN}=\Delta M  \quad .                \label{3.20}
\end{equation}
In order to get (\ref{3.20}), one finds that $k_1=\frac{M_0}{2}$, and therefore
the transformation law of $\Theta_{tt}$ is
\begin{equation}
     \delta_{\varepsilon(t)}\Theta_{tt}=\varepsilon(t)\Theta{'}_{tt}
+2 \Theta_{tt}   \varepsilon{'}(t)-
  \frac{M_0 r_0^2}{(n-1)^2}\varepsilon{'''}(t)   \quad .    \label{3.21}
\end{equation} 

\subsection{AdS$_2$-symmetries}

Alternatively we can consider the symmetries (\ref{3.6}-\ref{3.8}), which
preserve the asymptotic form of the two-dimensional metric. In this case the
invariant quantity under the ``pure gauge diffeomorphisms''
\begin{eqnarray}
      \zeta^t &=& \frac{\alpha^t(t)}{x^4}+\dots       \label{3.22}\\
   \zeta^x &=& \frac{\alpha^x(t)}{x}+\dots \quad ,     \label{3.23}
\end{eqnarray}
is
\begin{equation}
    \Theta_{tt}=k_2\left( \gamma_{tt}-\frac{8}{l^2}\gamma_{xx}\right)
                                                       \label{3.24}
\end{equation}
where $k_2$ is again a constant coefficient. The transformation law of
$\Theta_{tt}$ is
\begin{equation}
   \delta_{\varepsilon(t)}\Theta_{tt}=  \varepsilon(t)\Theta{'}_{tt}
  +2\Theta_{tt}\varepsilon{'}(t)-\frac{k_2l^2}{4}\varepsilon^{'''}(t)
            \quad .                                    \label{3.25}
\end{equation}
The value of $k_2$ can be determined immediately by requiring
\begin{equation}
        \Theta_{tt}\vert_{RN}=\Delta M\quad .         \label{3.26}
\end{equation}
We find that
\begin{equation}
       k_2=M_0                                         \label{3.27}
\end{equation}
and this reproduces the same transformation law as that of (\ref{3.21}).

\section{The role of the extra dimension and black hole entropy}

One of the effects of introducing the extra Kaluza-Klein coordinate is that
it converts the near-horizon geometry AdS$_2\times$S$^n$ (\ref{2.10}) into
the AdS$_3\times$S$^n$ geometry. This has important consequencies for the
causal structure of the boundaries. For AdS$_2$ the boundary consists of two 
disconnected time-like real lines. It can be embedded (more precisely, the
connected part corresponding to $x=+\infty$) into the cylindrical boundary 
of AdS$_3$ through the relations (\ref{2.17})-(\ref{2.18}). The lifted time-like
boundary of AdS$_2$ becomes a null slice of the AdS$_3$ boundary and then
one can perform the usual Fourier mode decomposition of $\Theta_{tt}(t)$,
which now bcomes $\Theta_{--}(x^-)$, as follows
\begin{equation}
       L^R_n = \frac{2Q\Delta_{(n)}^{1/2}}{2\pi}\int^{2\pi}_0 dx^-
     \Theta_{--}(x^-)e^{inx^-} \quad .                  \label{4.1}
\end{equation}
Using either (\ref{3.21}) or (\ref{3.25}) (with \ref{3.27}) we get a Virasoro 
algebra
($\varepsilon_m=e^{imx^-}$)
\begin{equation}
      \{ L^R_n,L^R_m\} =\delta_{\varepsilon_m}L^R_n=
  (n-m)L^R_{n+m}+\frac{c}{12}n^3\delta_{n,-m}          \label{4.2}
\end{equation}
with a central charge 
\begin{equation}
      c=\frac{24M_0r_0^2Q\Delta_{(n)}^{1/2}}{(n-1)^2}  \quad .\label{4.3}
\end{equation}
Note that for the $(n+2)$-dimensional theory (\ref{4.1}) involves a time
integration which is not permitted by the canonical formalism, only in the
higher-dimensional theory is that step allowed. Moreover the quantities
(\ref{4.1}) are conserved charges ({\it i.e.}, time independent). In
contrast, in the $(n+2)$-dimensional space-time only the zero mode
of $\Theta_{tt}$ is a well-defined conserved charge. This reinforces the 
idea that a higher-dimensional perspective is necessary to properly
define a Virasoro algebra leading, via Cardy's formula
\begin{equation}
 \log \rho\sim 2\pi \sqrt{\frac{cL_0^R}{6}}                           \label{4.4}
\end{equation} 
to the adequate density of states. The $L_0^R$-Virasoro charge takes the
value (for static R-N black holes)
\begin{equation}
      L_0^R=\frac{\Delta M}{2Q\Delta_{(n)}^{1/2}}       \label{4.5}
\end{equation}
and together with (\ref{4.3}) we obtain
\begin{equation}
     \Delta S=2\pi\sqrt{\frac{2r_0^2M_0}{(n-1)^2}\Delta M}   \label{4.6}
\end{equation}
in exact agreement with the entropy deviation from extremality of R-N black 
holes.
\section{Conclusions and final comments}

The search for a universal mechanism responsible for black hole entropy 
involves two basic ingredients: near-horizon symmetries and conformal field 
theory. For R-N black holes the near-horizon geometry contains AdS$_2$ and the
corresponding conformal generators live in the time-like boundary of AdS$_2$
thus making problematic the role of the Virasoro modes (in a time direction)
in classifying states. In the analogous situation in AdS$_3$, the space-like slice
of the boundary is a circle. The relevant diffeomorphisms are parametrized
by two functions which translate into two Virasoro algebras. The Virasoro 
generators are the modes on the space-like circle. Moreover, the holomorphicity 
of the chiral components of the stress tensor implies that one can convert
the integral over a spatial coordinate into an integral over a null coordinate.
Hence, it can be concluded that the Hilbert space on this circle has a growth
of states given by Cardy's formula. In this paper we have shown that the extra
Kaluza-Klein dimension allows to embed the time-like boundary of AdS$_2$
into a null slice of AdS$_3$ providing consistency to the use of Cardy's 
formula to evaluate the entropy deviation from extremality in a statistical
way.

\vskip .5cm
\noindent
{\bf Acknowledgements}. P. Navarro acknowledges the Ministerio de 
Educaci\'on y Cultura for a FPU fellowship. This work has been partially 
supported by the DGICYT (Spain), and by the Junta de Castilla y Le\'on (J.M.I.). 


\end{document}